\documentclass[traditabstract]{aa}

\usepackage{graphicx}
\usepackage{multirow}
\usepackage{natbib}
\usepackage{adjustbox}
\usepackage{txfonts}
\usepackage{lipsum} 
\usepackage{mathrsfs}
\usepackage{breqn}
\usepackage[colorlinks = true,linkcolor = blue,urlcolor = blue,citecolor = blue]{hyperref}
\usepackage{xcolor}
\definecolor{yellowgray}{rgb}{0.90, 0.90, 0.2}
\definecolor{bluegray}{rgb}{0.20, 0.60, 0.80}
\definecolor{palered}{rgb}{0.99, 0.40, 0.5}
\definecolor{darkgray}{rgb}{0.35, 0.35, 0.35}
\definecolor{darkgrayb}{rgb}{0.75, 0.75, 0.75}
\definecolor{palegray}{rgb}{0.96, 0.96, 0.96}

\begin{document} 

\title{Understanding the thermal and magnetic properties of an X-class flare in the low solar atmosphere}
  \titlerunning{Understanding the thermal and magnetic properties of a X-class flare in the low solar atmosphere}
   \author{F. Ferrente\inst{1}\and C. Quintero Noda\inst{2,3}\and
             F. Zuccarello\inst{1,4}\and S.L. Guglielmino\inst{4}}
   \institute{ Dipartimento di Fisica e Astronomia ``Ettore Majorana'', Università di Catania, Via S. Sofia 78, 95123 Catania, Italy\\
     \email{fabiana.ferrente@inaf.it}    
\and   Instituto de Astrof\'isica de Canarias, E-38205, La Laguna, Tenerife, Spain.   
\and  Departamento de Astrof\'isica, Univ. de La Laguna, La Laguna, Tenerife, E-38200, Spain
     \and INAF - Catania Astrophysical Observatory, Via S. Sofia 78, 95123 Catania, Italy         }
   \date{Received - February 2024 ; accepted - April 2024  }

 
\abstract{We analyse the spatial distribution and vertical stratification of the physical parameters of the solar atmosphere when an X-class flare occurs. We made use of observations acquired by the Interferometric Bidimensional Spectropolarimeter instrument when observing the full Stokes parameters for the \ion{Fe}{i} 6173~\AA \ and \ion{Ca}{ii} 8542~\AA \ transitions. We analysed the observed spectra using the newly developed DeSIRe code to infer the atmospheric parameters at photospheric and chromospheric layers over the entire observed field of view. Our findings reveal that the chromosphere is characterised by temperature enhancements and strong upflows in the flare ribbon area, which indicates that the flaring event is producing hot material that is moving outwards from the Sun. We did not detect any trace of temperature enhancements or strong velocities (of any sign) at photospheric layers, signalling that the impact of the flaring event mainly happens at the middle and upper layers. The information about the magnetic field vector revealed relatively smooth stratifications with height for both magnetic field strength and inclination. Still, when examining the spatial distribution of the magnetic field inclination, we observed the presence of large-scale mixed polarities in the regions where the flare ribbon is located. These results suggest that the interaction between those mixed polarities could be the flare's triggering mechanism.}

\keywords{Sun: flares, magnetic fields, chromosphere -- Techniques: polarimetric, high angular resolution --  Radiative transfer}

\maketitle

\section{Introduction}

Solar flares are one of the most powerful magnetic events in the Solar System. They are sudden energy releases within active regions that emit vast amounts of radiation across the electromagnetic spectrum, from radio to gamma rays, typically within a few minutes to tens of minutes, depending on the flare phase. Solar flares can have different sizes, and their main classification is based on the soft X-ray flux at 1-8 $\AA$ measured by the Geostationary Orbiting Environmental Satellites (GOES). Flares are classified into X, M, C, B, and A flares, with X being the most potent. They develop in two main phases: the impulsive and the gradual phase. The impulsive phase is the most intense part of a flare and is where hard X-rays, gamma rays, microwaves, and white light are emitted. These signals show that electrons and ions are accelerated to high speeds, reaching velocities typically in the range of a few tens of to a thousand kilometres per second. In this phase, the particles carry the energy to the lower layers of the Sun, where they heat the gas and make it glow brightly in different wavelengths. The places where the particles hit the Sun’s surface are called footpoints or ribbons. They mark the thick ends of the magnetic loops, where most of the flare energy is converted to radiation from the chromosphere and photosphere \citep[e.g.,][]{1974hirayama,2011fletcher}.

The lower solar atmosphere therefore plays a crucial role in our understanding of solar flares. Some authors have used spectropolarimetric data to infer the physical processes involved in these events. Particularly, they used numerical codes, so-called inversion codes, that solved the radiative transfer equation iteratively until the observed profiles were fitted with a high accuracy. For instance, \cite{2021yadav}, examined a C2 class flare through the inversion of the \ion{Ca}{ii} K, \ion{Ca}{ii} 8542~$\AA$, and \ion{Fe}{i} 6173~$\AA$ transitions using the STockholm inversion Code \citep[STIC;][]{stic}. Their findings highlighted the presence of bright footpoints and atmospheric heating up to 11 kK in the upper layers. They also suggested that the \ion{Ca}{ii} spectral lines are sensitive to deeper layers (i.e. $\log\tau$ = -3 instead of $\log\tau$ =-4), leading to an apparent increase in the line-of-sight (LOS) magnetic field. \cite{2017libbrecht} investigated the neutral helium triplet line \ion{He}{i} D3 at 5876~$\AA$ during a C3.6 class flare, employing the HAZEL \citep{hazel} inversion code. They also detected the presence of chromospheric condensation, manifested as downflows of about 50 km/s in cool chromospheric lines and the redshifted component of \ion{He}{i} D3. 
Moving to higher class flares, \cite{2017Kuridze} analysed a C8.4 class flare, focusing on the \ion{Ca}{ii} 8542~$\AA$ spectral line. The authors used the NICOLE \citep{nicole} code to study the temperature and velocity evolution in the flaring chromosphere, revealing intense heating at the footpoints during the flare peak. In a subsequent study, \cite{2018Kuridze} investigated the temperature and magnetic field structure and evolution of an M1.9 class flare once again using NICOLE to invert the \ion{Ca}{ii} 8542~$\AA$ transition. The comparison between flaring and non-flaring regions showed intense heating of the flare ribbon during its peak and increased sensitivity of \ion{Ca}{ii} to the lower atmosphere, where the magnetic field was stronger.

Additionally, \cite{2015kuckein} conducted a detailed analysis of an M3.2 solar flare. This research focused on the \ion{He}{i} 10830~$\AA$ spectral region and captured the full Stokes spectropolarimetric data during the pre-flare, flare, and post-flare phases. Key observations included the strong emission of the \ion{He}{i} 10830~$\AA$ intensity during the flare, which was contrasted by the nearby \ion{Si}{i} 10827~$\AA$ spectral line in the photosphere that remained in absorption. The study also highlighted significant changes in the photospheric magnetic field, with detections of magnetic field concentrations of up to 1500~G before the flare, which subsequently decreased.
Also, \cite{2021vissers} studied a confined X2.2 solar flare and conducted separate inversions for different spectral lines, specifically both local thermodynamic equilibrium (LTE) and non-LTE (NLTE) inversions of the \ion{Fe}{i} 6301.5~$\AA$ and 6302.5~$\AA$ transitions and spatially regularised weak-field approximation (WFA) and NLTE inversions of the \ion{Ca}{ii} 8542~$\AA$ spectral line. This approach allowed them to analyse the magnetic field vector in the photosphere and chromosphere. The results highlighted strong correlations between photospheric fields inferred in LTE and NLTE conditions and a good alignment between the NLTE and WFA-inferred chromospheric fields.
 
We analysed in \cite{Ferrente2023} an observation of an X-class flare, focusing mainly on the properties and evolution of the longitudinal component of the magnetic field during the flare and post-flare phases. On this current occasion, we are interested in expanding that work using the same observation but analysing the spatial and vertical distribution of the thermal and magnetic field parameters for the entire observed field of view (FOV). The analysis consists of performing NLTE inversions of a multi-line full Stokes observation of an X-class flare.

The main goal of this study is to examine the stratification of the atmospheric parameters related to this energetic X-class flare, focusing on both flaring and non-flaring regions across the entire FOV. Our method involves an in-depth analysis of the full Stokes parameters, analysing the profiles of the \ion{Fe}{i} 6173 Å and \ion{Ca}{ii} 8542 Å transitions simultaneously, as well as the spatial distribution and vertical stratification of the inferred atmospheric parameters.

\section{Data and methodology}\label{Method}

\subsection{Data}

We study in this work the properties of AR12192 using observations acquired by the Interferometric BIdimensional Spectropolarimeter \citep[IBIS;][]{Cavallini2006} at the Dunn Solar Telescope \citep[DST,][]{Dunn1969}. The data set used in this publication was recorded on 2014 October~22 from 14:29~UT to 15:41~UT. During this period, an X1.6 flare took place. The instrument recorded the full Stokes vector for the \ion{Fe}{i} 617.3~nm and \ion{Ca}{ii} 854.2~nm transitions and the intensity information for the H$\alpha$ 656.29~nm line. The time needed to observe the three spectral lines, one after the other, was 51~s, while the time difference between the \ion{Fe}{i} 617.3~nm and \ion{Ca}{ii} 854.2~nm scan is 31~s. The scan was repeated 82 times to record the evolution of the active region. This data set is also available in the IBIS archive \citep[IBIS-A,][]{ibis_a}.

In this work, we are mainly interested in analysing the polarimetric information of the flare, so we focused on the \ion{Fe}{i} and \ion{Ca}{ii} spectral lines only, leaving the analysis of the Balmer transition for future works. Moreover, as it is the first time (as far as we know) that someone has attempted to fit spectropolarimetric observations of an X-class flare of these two lines simultaneously with an inversion code, we focus on the first snapshot of the time series. This is because part of the work is dedicated to finding the best approach for inverting these observations and studying the accuracy and limitations of the method used. Later in a following publication, we will examine the evolution of the atmospheric parameters during the flare activity.

\begin{figure*}
	\begin{center} 
		\includegraphics[trim=0 0 0 0,width=18cm]{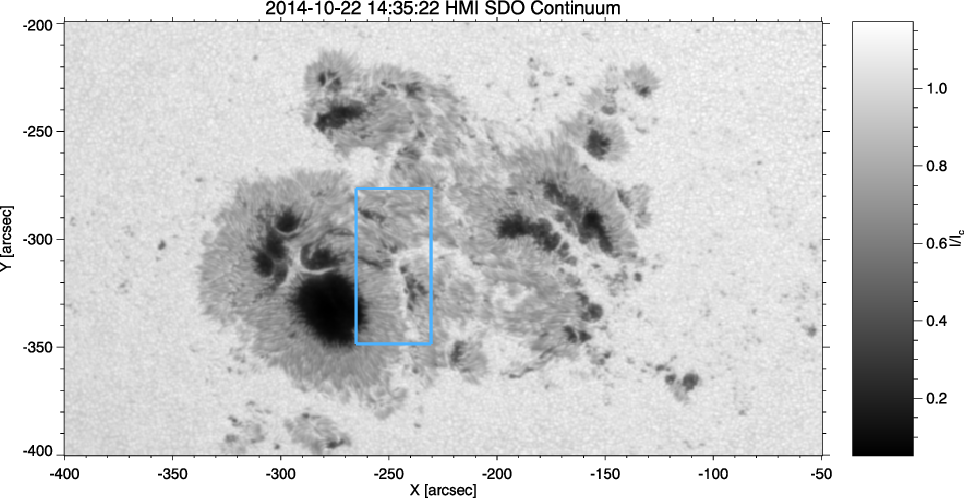}
		\vspace{-0.10cm}
		\caption{SDO/HMI continuum image of the AR12192. The blue rectangle corresponds to the IBIS observed FOV.}
		\label{continuum}
	\end{center}
\end{figure*}

As a reference, we show in Figure~\ref{continuum} the spatial distribution of the continuum intensity at \ion{Fe}{i} 617.30~nm recorded by the Helioseismic and Magnetic Imager \citep[HMI;][]{Scherrer2012} on board the Solar Dynamics Observatory \citep[SDO;][]{Pesnell2012}. We use these data to depict the large size and complexity of AR12192 and to highlight (see blue square) the IBIS observed FOV. The instrument was centred on the region where the flare ribbon was located, at approximately (-250,-300) arcsec. We study the complete FOV observed by IBIS in this work. 

\subsection{Methodology}

This work aims to understand the properties of the atmospheric parameters where the X-class flare took place. In particular, we want to infer the properties of the atmosphere at different heights in the photosphere and chromosphere using the information provided by the \ion{Fe}{i} and \ion{Ca}{ii} spectral lines. In order to infer the atmospheric parameters, we used the Departure coefficient aided Stokes Inversion based on Response functions \citep[DeSIRe;][]{DeSIRe} code for analysis of the observations. The code can perform inversions under LTE and NLTE for any number of atomic species. In our case, we solved the atomic populations for the \ion{Fe}{i} 617.30~nm transition under the LTE assumption, while we considered NLTE conditions for the case of the \ion{Ca}{ii} 854.2~nm spectral line. We used a five-level plus continuum model for \ion{Ca}{ii}, which is included in the default DeSIRe library, and it is based on the work of \cite{Shine1974}. The \ion{Ca}{ii} transition was treated under complete redistribution, which seems to be a reasonable approach for the three infrared transitions of \ion{Ca}{ii} \citep[e.g.,][]{Uitenbroek2006}. Inversions were performed under the plane-parallel approximation, and the solution of each pixel is independent of the others. Abundance values are those provided by \cite{Asplund2009}.

In order to get the data ready to be used with DeSIRe, we first normalised the Stokes profiles for each spectral line, computing the average continuum wavelength intensity over a quiet Sun region. In the case of the \ion{Ca}{ii} 854.2~nm, the local continuum was not reached. That is, what we consider to be the continuum for the \ion{Ca}{ii} corresponds to the furthest wavelength point at  853.99~nm, which falls in the blue wing of the broad infrared spectral line. In this case, we followed a two-step process to normalise the data. We first normalised the Stokes profiles using the averaged quiet Sun continuum intensity at this wavelength (i.e. 853.98~nm), which is similar to what we would do if it were an actual continuum wavelength point. Then, we performed a synthesis with DeSIRe using the 1D semi-empirical HSRA atmosphere \citep{1971HSRA}, including the shift on optical depths due to the heliocentric angle of our observation (i.e.  $\mu = 0.9$). The HSRA atmosphere is used internally in DeSIRe to compute the local continuum at any wavelength, so we can derive from the synthetic profile the intensity scaling factor we need to apply to re-normalise the Stokes profiles as if we had actually observed the average quiet Sun local continuum.

As this was the first time we have planned to use the DeSIRe code to fit spectropolarimetric data of an X-class flare, we decided to take a multi-step approach. First, we fit the Stokes profiles for the \ion{Fe}{i} 617.30~nm transition under the LTE approximation, studying different inversion configurations until a satisfactory fit was obtained. After that, we repeated the same process, fitting only the \ion{Ca}{ii} 854.2~nm spectral line. In this case, we worked under NLTE approximation and performed multiple tests until the fit of the Stokes vector was accurate. Finally, we inverted almost simultaneous data from the two transitions together, starting from the configuration obtained for the \ion{Ca}{ii} (the transition that shows the most demanding profiles, as we explain later), and we continued testing various configurations until a good fit of both transitions was obtained.

The final inversion configuration for the nodes is shown in Table~\ref{tab:Config1}. As a reminder, the inversion code uses as free parameters the values of the physical quantities on a specific grid of optical depth points (nodes), taking into account that the inferred atmosphere will be the result of an interpolation between those nodes (except in the case of one or two nodes that will be a constant or linear perturbation). Regarding the physical quantities, we inferred the temperature, LOS velocity, microturbulence, magnetic field vector, and macroturbulence during the inversion process. No information regarding the instrument's point spread function was used. The inversion was done in multiple cycles, increasing the number of nodes for each atmospheric parameter as presented in Table~\ref{tab:Config1}, except for the macroturbulence.

The inversion was done in parallel using the Python-based wrapper described in \cite{Gafeira2021}. This code allows the distribution of the pixels to be inverted among the CPU threads available on a computer. In addition, the wrapper can provide a set of initial atmospheres so each pixel can be fitted by starting each inversion from a different initial atmosphere and then picking the solution that provides the best $\chi^2$ value. This method allows the presence of the typical salt-and-pepper pattern that can appear on specific atmospheric parameters to be reduced, particularly the LOS velocity and the magnetic field inclination. Moreover, we could define a threshold for the minimum $\chi^2$ value we wanted to achieve to speed up the process. If that value is obtained using, for example, the first atmosphere, the code will move to the next pixel instead of repeating the inversion multiple times. In our case, we used a final set of 21 initial atmospheres and a threshold that, while yielding accurate results, took an equivalent time inverting each pixel five times, saving 78\% of the time we would need to invert each pixel 21 times. 

Regarding the specific atmospheres used in the inversion as initialisation, they were created from the FAL models presented in \cite{1993fontenla} and the HSRA atmosphere by adding small amplitude variations in the LOS velocity, microturbulence, and magnetic field (which are null in the original models). We created a library of initial atmospheres with constant values with height for those physical parameters. Moreover, we chose specific configurations, for example, an LOS velocity could be positive or negative; the magnetic field could be weak or strong; the inclination could be below 90 or above 90 degrees. These combinations provided stark changes on the profiles generated by the code (e.g. a different Stokes $V$ polarity or redshifted and blueshifted Stokes profiles). Also, after some tests, we refined the initial set of atmospheres, adding some atmospheres obtained during the inversion as starting models. In that sense, we improved the overall results by getting a smooth spatial variation (low presence of salt-and-pepper) for the atmospheric parameters.

\begin{center}
	\begin{table}
		\vspace{+0.4cm}
		\caption{Number of nodes used for the inversion of each atmospheric parameter. Each new column towards the rightmost side of the table adds (or maintains) more nodes for the inversion and corresponds to an additional inversion cycle.} 	
		\begin{adjustbox}{width=0.45\textwidth}
			\bgroup
			\def\arraystretch{1.25}
			\begin{tabular}{|l|cccc|}
				\hline
				\multirow{2}{*}{\textbf{Parameter}}     & 
				\multicolumn{4}{c|}{\textbf{Nodes}}   \\
				& \textbf{Cycle 1} & \textbf{Cycle 2} &  \textbf{Cycle 3} & \textbf{Cycle 4}  \\
				\hline
				Temperature & 2 & 3 & 5 & 6 \\
				Line-of-Sight Velocity & 1 & 2 & 3 & 5 \\
				Field Strength & 1 & 2 & 3 & 3 \\
				Inclination & 1 & 2 & 3 &3 \\
				Azimuth & 1 & 1 & 1 & 1 \\
				Microturbulence & 1 & 1 & 2 & 2 \\
				\hline                       
			\end{tabular}
			\egroup
		\end{adjustbox}
		\label{tab:Config1}    
	\end{table}
\end{center}

\begin{figure*}
	\begin{center} 
		\includegraphics[trim=0 0 00 0,width=18cm]{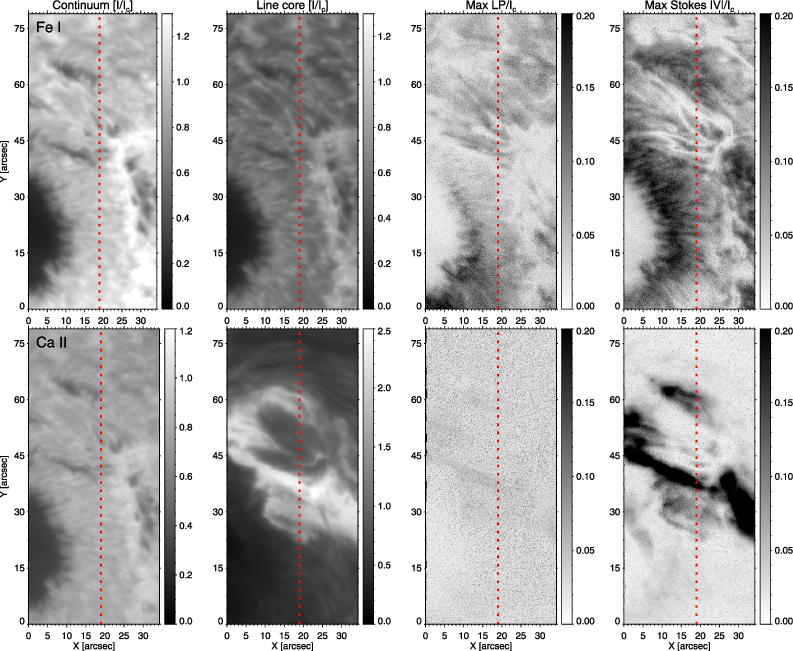}
		\vspace{-0.10cm}
		\caption{Spatial variation of various spectral features. From left to right, we show the continuum intensity, line core intensity, and maximum linear and circular polarisation signals. The top row corresponds to the \ion{Fe}{i} 617.30~nm transition, while the bottom row shows the results for the \ion{Ca}{ii} 854.2~nm spectral line. The dotted vertical line highlights a region we examine later on. For the \ion{Fe}{i} transition, the continuum wavelength is measured at 617.31~nm and the line core at 617.33~nm. Similarly, for the \ion{Ca}{ii} transition, the continuum and line core are measured at 853.98~nm  and 854.22~nm, respectively.}
		\label{fig:obsprof}
	\end{center}
\end{figure*}

\section{Results}

\subsection{Spectral information}

\subsubsection{Spatial distribution of polarisation signals}

We started the analysis of the spectral features of the region where the X-flare event takes place by examining several general quantities. In particular, we studied the spatial distribution of the continuum intensity, line core intensity, and maximum linear and circular polarisation signals for both transitions. Regarding the maximum linear polarisation signal, we computed it as the maximum value for the observed spectral range of $\sqrt{Q^2+U^2}$. The maximum circular polarisation corresponds directly to the maximum signal of the absolute value of Stokes $V$ over the entire observed wavelength range for each spectral line. We show the results for the two transitions in Figure~\ref{fig:obsprof}.

	\begin{figure*}
		\begin{center} 
			\includegraphics[trim=0 0 0 0,width=18.3cm]{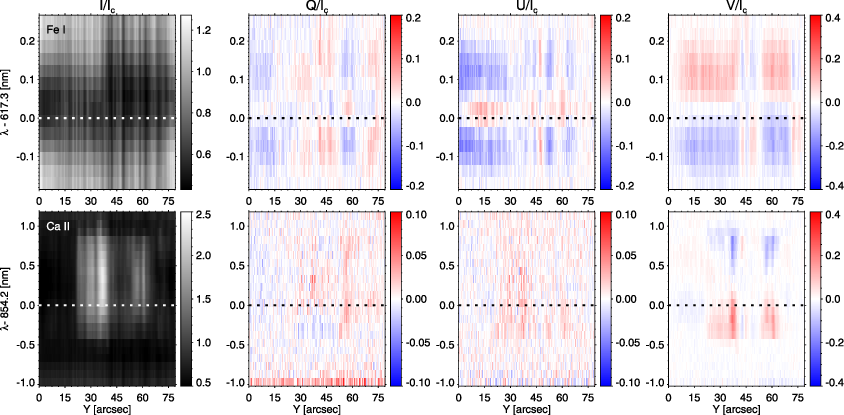}
			\vspace{-0.10cm}
			\caption{Maps corresponding to a vertical cut of the observed Stokes profiles. The abscissa axis represents the region highlighted by the vertical dotted line in Figure~\ref{fig:obsprof}, while the ordinate axis corresponds to the spectral domain for the \ion{Fe}{i} (top) and \ion{Ca}{ii} (bottom) transitions, respectively.}
			\label{fig:obscut}
		\end{center}
	\end{figure*}

Starting with the continuum intensity at 617.3~nm, we could see part of a sunspot in the bottom-left corner of the FOV and various complex structures in the rest of the FOV. Multiple small-scale areas have low intensity values (dark in this image), probably corresponding to regions of enhanced magnetic field strength. We could also detect areas with high intensity values on the right side of the observed FOV. The mentioned distribution of intensity signals is also visible at the \ion{Fe}{i} line core wavelengths, with no stark differences except the overall drop of intensity values. When we examined the maximum linear polarisation signals, we could trace the presence of the penumbra associated with the sunspot located at the left side (outside of the FOV of IBIS; see also Figure~\ref{continuum}). Thus, large amplitude linear polarisation signals exist in many areas with some structure around the FOV contained on the horizontal band defined between $Y=[40,60]$~arcsec. These areas also show a particular distribution of signals for the Stokes $V$ parameter, with a complex thread-like pattern, and the areas with strong linear polarisation signals seem to correspond to weak Stokes~$V$ signals, and vice versa. This may be due to the presence of a complex magnetic field distribution that changes abruptly in short spatial scales due to, for example, a twisted field line configuration.

Moving to the \ion{Ca}{ii} transition (see the bottom row in Figure~\ref{fig:obsprof}), what we defined as a continuum resembles the first two leftmost panels of the upper row. The contrast is not as high as in the continuum at 617.3~nm because, as mentioned before, the furthest spectral position observed in the 854.2~nm channel does not reach the continuum. Moving towards the core of the spectral line, we found a different spatial distribution, as those wavelengths are mainly sensitive to the chromosphere. We could see a stark enhancement of the line core intensity with values higher than the estimated average local quiet Sun continuum at 854.2~nm, indicating that the line core is in emission. These intensity enhancements correspond to the ribbon of the X-flare event that occurred during the observation. They are located mainly on a narrow lane around the centre of the FOV, between the sunspot on the left side and the complex structures on the right side (see also Figure~\ref{continuum}). Regarding the linear polarisation signals, we did not detect a clear presence above the noise. It seems there are signals in the region of highest intensity in the line core intensity panel, but they are faint and have amplitudes close to the noise level. However, in the case of the Stokes $V$ parameter, the situation is different, and we have a direct correlation between the high intensity values for the line core and the presence of large amplitude circular polarisation signals (see rightmost bottom panel). Thus, the region with high line core emission seems co-spatial, with almost longitudinal (linear polarisation signals are weak) magnetic fields in the chromosphere. We note that contrary to the photospheric \ion{Fe}{i} 617.3~nm transition, in the case of the \ion{Ca}{ii} transition, the circular polarisation signals on the ribbon areas dominate the observed FOV, as the signals produced at the sunspot in the bottom-left side of the FOV are much weaker in comparison.

\subsubsection{Spectral information over a 1D cut}
	
We wanted to explore further the spectral properties of the flare ribbon area and the surrounding region by examining the entire observed spectral range for multiple spatial locations. In this regard, we show in Figure~\ref{fig:obscut} the evolution of the four Stokes parameters over the vertical line shown in Figure~\ref{fig:obsprof}. In the case of the photospheric \ion{Fe}{i} 617.3~nm transition (top row), we have that the Stokes $I$ profile shows intensity variations on the spatial domain that take place over small scales, around a few arc seconds. The reason for these high-frequency spatial variations is the presence of various features on the observed FOV, including the granulation. Regarding the spectral domain, wavelengths close to the line core show a darkening between [30 and 50]~arcsec, roughly where the flare ribbon was observed. In the case of Stokes $Q$ and $U$, there are polarisation signals of more than 10\% of the averaged quiet Sun continuum intensity. Both Stokes parameters display significant amplitude variations along the selected FOV. Moreover, the sign of the Zeeman components also fluctuates over the selected FOV, with more prominent sign changes between [30 and 50]~arcsec, that is, around the area where we detected the presence of the flare ribbon. In the case of Stokes $V$, we have amplitude variations on small spatial scales again, although the Stokes $V$ polarity generally remains the same for the selected FOV. We detected a change of polarity only around the ribbon area, roughly between [30 and 50]~arcsec, although the amplitude of the signal in those areas is similar to the rest of the regions.
	
The spectral information of the chromospheric \ion{Ca}{ii} transition has a starker contrast, with a significant intensity enhancement that reaches levels higher than two of $I$/$I_c$ in the central part of the FOV, mainly between [30 and 50]~arcsec. Those intensity enhancements primarily happen at line core wavelengths. At the same time, the continuum seems to remain at levels similar to that of the areas outside the ribbon region. In this case, linear polarisation profiles seem to be dominated in general by the noise of the observation. However, we could detect the presence of faint (in comparison with the floor noise level) linear polarisation signals in the region where the flare ribbon was observed. In the case of the Stokes $V$ parameter, we have high amplitude signals mainly for the areas associated with the flare and for wavelengths close to the line core. We also detected multiple changes of polarity where the ribbon appeared (i.e. around [40 and 50]~arcsec).

\subsection{Inversion results}
	
\subsubsection{Accuracy of the profiles}
	
\begin{figure}
		\centering
		\includegraphics[width=8.5cm]{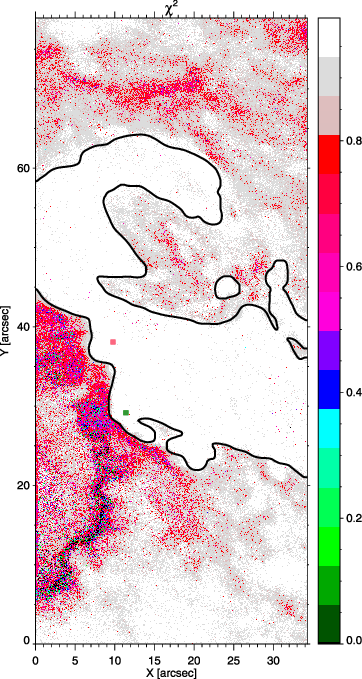}
		\caption{Spatial distribution of the inverse $\chi^2$ over the IBIS observed FOV. Values corresponding to one represent a highly accurate fit, while values close to zero correspond to a poor fit. The green and pink squares designate the location of reference pixels we study later. The black contour highlights the location of the flare ribbon, and it corresponds to the areas with enhanced intensity on the line core of the \ion{Ca}{ii} transition (see Figure~\ref{fig:obsprof}).}
		\label{fig:chi}
	\end{figure}

We made use of the DeSIRe inversion code for the fitting of the full Stokes spectra of the \ion{Fe}{i} 617.3~nm  and \ion{Ca}{ii} 854.2~nm transitions simultaneously. As this was the first time we have used the code to fit these lines in such a complex scenario (i.e. a flare eruption), we ran multiple inversion tests with various configurations until the $\chi^2$ maps provided reasonable results over the observed FOV. We present the results of the merit function for the best inversion configuration we obtained in Figure~\ref{fig:chi}. The figure displays the inverse normalised $\chi^2$, where a perfect fit corresponds to one and a poor fit is close to zero.
	
We observed that the values are between 0.95 and 1 in the area where the flare ribbon is located, indicating that the code fits the observed profiles accurately. These results are encouraging because this area is characterised by very complex profiles (i.e. the intensity profile of the \ion{Ca}{ii} line is in emission). Considering the rest of the observed FOV, we observed that the overall distribution suggests a good fit accuracy, as the $\chi^2$ is mainly between 0.75 and 1. The only region where the code struggles to produce good fits is close to the umbra and penumbra of the sunspot located in the bottom-left corner. We believe one reason for the low accuracy in that area is the significant differences in terms of physical properties with respect to the flare region. There, the plasma is cooler and moves at low speeds, so maybe our configuration (in terms of initial atmospheres, weights for the Stokes parameters, and nodes for the atmospheric parameters) is not best suited to that area with such a different physical scenario. We will explore this issue in future works. However, as that region is not the main target of the present publication, we opted to use the current inversion configuration.

	\begin{figure*}
		\begin{center} 
			\includegraphics[trim=0 0 0 0,width=18cm]{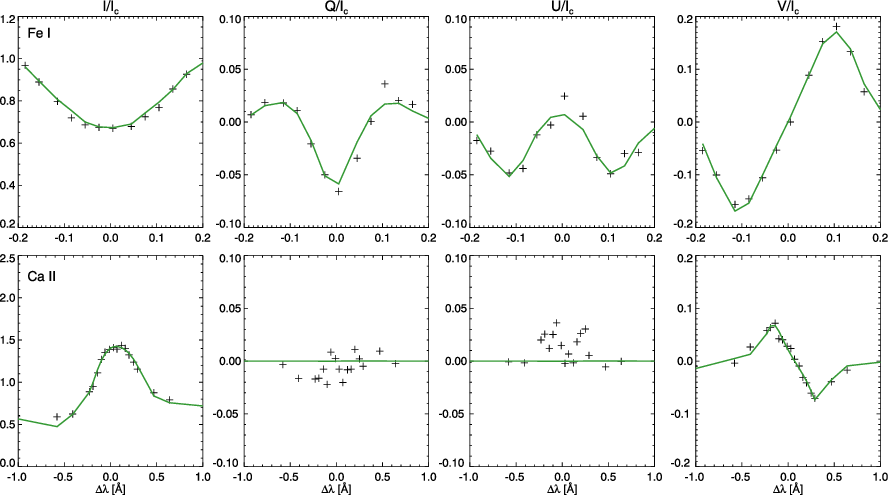}
			\vspace{-0.10cm}
			\caption{Comparison between the observed and the fitted Stokes profiles. The top and bottom rows show the Stokes profiles for the \ion{Fe}{i} and \ion{Ca}{ii} transitions, respectively. From left to right, we have the Stokes~$I$, $Q$, $U$, and $V$ parameters, respectively. Crosses correspond to the observed profiles, while solid coloured lines represent the fit obtained during the inversions. The spatial location of this pixel is highlighted in green in Figure~\ref{fig:chi}.}
			\label{fig:profilesgreen}
		\end{center}
	\end{figure*}
	
	\begin{figure*}
		\begin{center} 
			\includegraphics[trim=0 0 0 0,width=18cm]{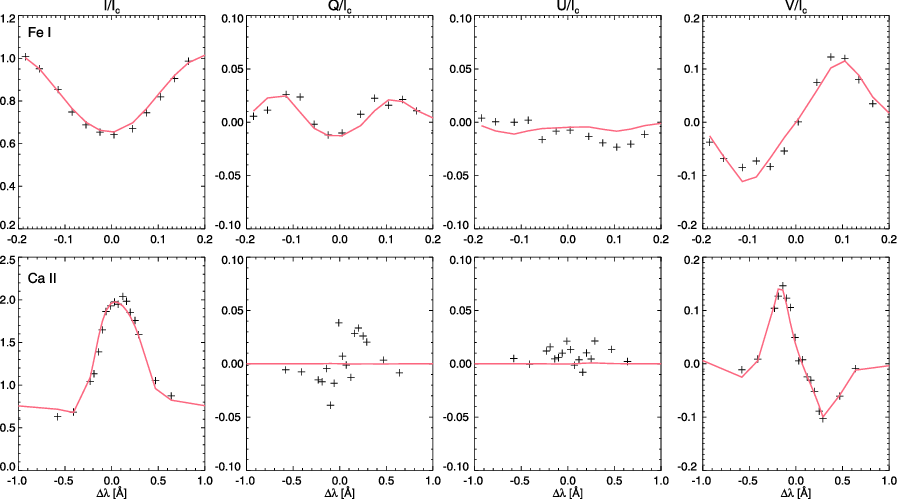}
			\vspace{-0.10cm}
			\caption{Same as Figure~\ref{fig:profilespink} but for the spatial location highlighted in pink in Figure~\ref{fig:chi}.}
			\label{fig:profilespink}
		\end{center}
	\end{figure*}

To check the quality of the fits in more detail, we also examined two specific profiles located in different areas along the flare ribbon. We selected those pixels based on two simple criteria: They are very complex, and the fits represent the overall accuracy we found when inverting the complete map. The location of the pixels on the observed FOV is highlighted with the coloured squares in Figure~\ref{fig:chi}. The corresponding profiles are displayed in two separate figures, Figure~\ref{fig:profilesgreen} and \ref{fig:profilespink}. The configuration of both figures is identical, with the Stokes profiles for the \ion{Fe}{i} transition in the top row and those corresponding to the \ion{Ca}{ii} spectral line presented in the bottom row. In each figure, we plot the original profiles with crosses and the results from the inversion with solid lines, whose colours match that of the squares plotted in Figure~\ref{fig:chi}.

In the case of the first pixel (i.e. green square in Figure~\ref{fig:chi}), the intensity profiles (first column) are fitted with high accuracy. This is remarkable in the case of \ion{Ca}{ii}, which appears in emission with intensity values up to 1.5 of $I_c$. The linear polarisation signals for the \ion{Fe}{i} are easily detectable above the noise and well-fitted in general. In the case of the \ion{Ca}{ii}, the fits are poor, with no match between the observed signals and the inferred profiles. Finally, in the case of Stokes~$V$, we obtained again a high accuracy on the fit of both spectral lines. The situation is similar for the second pixel selected. We added this one because it is an even more extreme case than the previous one, with intensity signals for the line core of \ion{Ca}{ii} that go up to 2.0 of $I_c$. Here again, the fits are accurate for both lines for intensity and circular polarisation. The linear polarisation is fitted with reasonable accuracy for the \ion{Fe}{i} transition, but again, the fits are poor for the \ion{Ca}{ii}. We do not yet know precisely why there is a lack of accuracy for the linear polarisation signals of the chromospheric line, as the fits are accurate for the rest of the parameters. Maybe the issue can be attributed to the distribution of weights used; the highly intricate nature of these profiles, which likely present physical conditions that are challenging for our inversion configuration; or any other problem. We plan to further investigate this point in the future, but for now we believe the presented profiles are an excellent example of our accuracy in general, an accuracy that we think is high enough to analyse the atmospheric parameters inferred by the inversion code.

\subsubsection{Atmospheres}

\begin{figure*}
	\begin{center} 
		\includegraphics[trim=0 0 0 0,width=11cm]{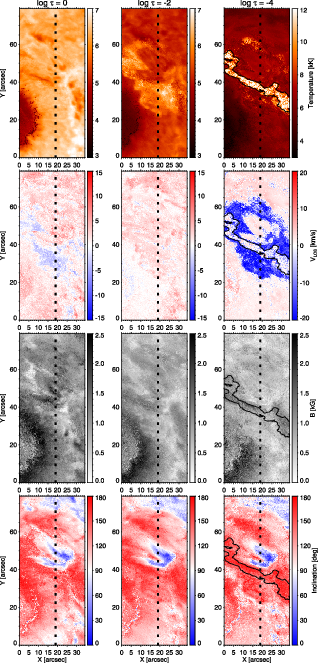}
		\vspace{-0.10cm}
		\caption{Variation with height of the atmospheric parameters across different optical depths in the solar atmosphere. Columns represent three optical depths, from left to right,  $\log \tau=[0, -2, -4]$. Rows display, from top to bottom, the temperature, LOS velocity, magnetic field strength, and inclination. The dashed line highlights a region crossing the flare ribbon, which we study later. We also added, as a reference, a contour line representing the area of temperature enhancement at  $\log \tau=-4$ in the panels of the rest of the atmospheric parameters at the same optical depth.}
		\label{fig:atmospheric_param}
	\end{center}
\end{figure*}

We examined the spatial distribution of the inferred atmospheric parameters at three selected optical depth values: $\log {\tau}= [0, -2, -4]$, where $\tau$ refers to the optical depth at 500~nm. Those optical depths roughly correspond to the lower and middle photosphere and lower chromosphere. Figure~\ref{fig:atmospheric_param} shows the results for the temperature, LOS velocity, magnetic field strength, and inclination.

Starting with the temperature at $\log {\tau}=0$, we observed a good correlation with the continuum intensity presented in Figure~\ref{fig:obsprof}. Areas with low intensity, such as the sunspot umbra in the bottom-left corner, appear as cool regions in the temperature panel, with values around 3000 to 4000 K. At the same time, the bright areas in the right part of the intensity maps related to the flare activity appear as hot areas in the temperature map, with values between 7000 and 9000 K. The rest of the regions show a pattern similar to the intensity panel, with temperature values between the two extreme cases mentioned before. When we examined the higher layers (i.e. $\log {\tau}=-2$), we saw a different landscape dominated by a central area that is hotter than its surroundings, with temperatures up to 10000 K. This hotter area corresponds to the flare ribbon. There is a direct correlation between the spatial distribution of temperature at this layer and the intensity signals on the line core of the \ion{Ca}{ii} infrared line (see bottom row and second panel from the left in Figure~\ref{fig:obsprof}). Moving higher up in the atmosphere, at $\log {\tau}=-4$, we have a different spatial distribution for the temperature. There is a thin structure diagonally crossing the middle of the observed FOV that is hotter than its surroundings, with temperature values reaching up to 15000 K, corresponding to more than three times the average temperature of its surroundings. There is no apparent trace of this specific feature in the intensity signals from the \ion{Fe}{i} nor \ion{Ca}{ii} transitions. However, there seems to be a close relation with the spatial distribution of circular polarisation signals for the \ion{Ca}{ii} spectral line.   

Examining the LOS velocity (second row of Figure~\ref{fig:atmospheric_param}) at $\log {\tau}=0$ revealed a complex velocity pattern with multiple changes of its sign over the entire FOV. However, we detected the presence of a patch of upward velocities (blue colour) in the middle of the observed region that is co-spatial with the temperature enhancement seen at $\log {\tau}=-2$. This patch is easy to detect, as it is surrounded by low velocities that change their sign as one moves further from this area. When studying the results at  $\log {\tau}=-2$, we did not see any specific pattern; only the sunspot shows velocities close to zero, while the rest of the FOV is, on average, dominated by weak downward motions (red colour). However, when we examined the highest optical depth selected for this comparison (i.e. $\log {\tau}=-4$), we found a stark contrast between the region where the flare activity occurs and its surroundings. There, we observed a complex feature dominated by strong upward motions up to -20 km/s and very low downward velocities (around 1-2 km/s). Interestingly, there is a clear correlation between the enhanced temperature region seen at $\log {\tau}=-4$ and a thin thread of low velocities surrounded by strong blueshifted motions at $\log {\tau}=-4$ as well. Besides that, there is also another region almost void of velocities in the middle of the complex feature that looks like two protrusions and seems to be correlated with the area with low temperatures seen on the temperature panel at $\log {\tau}=-2$ (around [20,50]~arcsec) and can also be seen as a low-intensity patch in the line core intensity signals of \ion{Ca}{ii} presented in Figure~\ref{fig:obsprof}.

The third panel of Figure~\ref{fig:atmospheric_param} presents the variation in the magnetic field strength across different atmospheric heights, as indicated by the $\log \tau$ values. The photospheric layer ($\log \tau = 0$) has the strongest magnetic field (darker colour) in the whole FOV, with values up to 1.4 kG. The magnetic field becomes weaker (brighter colour) as the height increases throughout most of the FOV. However, the region corresponding to the flare ribbon maintains its magnetic field strength, with values between 1.2 and 1.4 kG, even at higher layers of the atmosphere.

The last row of Figure~\ref{fig:atmospheric_param} displays the magnetic field inclination. One can see that the mentioned complex double protrusion feature (at around [20,50]~arcsec) appears at all atmospheric layers. It has a single polarity surrounded by a region of nearly horizontal magnetic fields ($\gamma\sim90$) and then by a magnetic field of opposite polarity. The transition of polarity change between both areas is smooth (see white-coloured areas), mainly at the middle and upper layers.

\begin{figure*}
	\centering
	\includegraphics[trim=0 0 0 0,width=18.8cm]{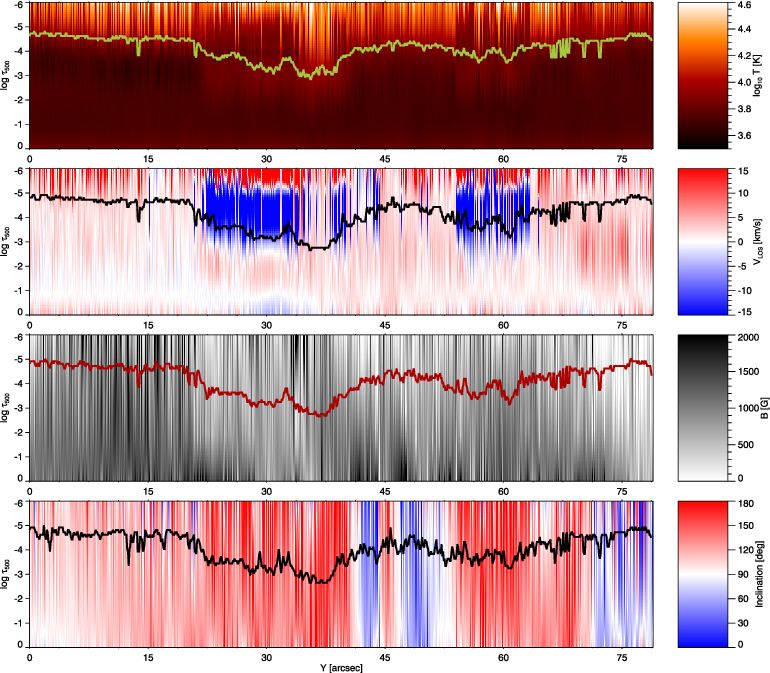}
	\vspace{-0.3cm}
	\caption{Height stratification of the atmospheric parameters for the region highlighted with a dashed line in Figure~\ref{fig:atmospheric_param}. Each panel shows, from top to bottom, the temperature, LOS velocity, magnetic field strength, and inclination, respectively. Inside each panel, there is a solid line that corresponds to the height where the sensitivity to changes in the atmospheric parameters is highest for the \ion{Ca}{ii} (more information in Sec.~\ref{RF_acc}).}
	\label{fig:cutx200arc}
\end{figure*}

We wanted to study the properties of the atmospheric parameters further, in particular, their evolution with height on short spatial scales. In this regard, we show in Figure~\ref{fig:cutx200arc} the vertical stratification of temperature, LOS velocity, magnetic field strength, and inclination along the region highlighted in Figure~\ref{fig:atmospheric_param} with a dashed line. Thus, in this case, the 2D representation corresponds to a fixed spatial location along the FOV (i.e. in the $X$-axis) and represents the evolution with optical depth of the atmospheric parameters along the $Y$-axis.

Starting with the inferred temperature (see top panel of Figure~\ref{fig:cutx200arc}), we observed significant variations of its amplitude in the optical depth range used for the inversion, so we opted to use a colour bar that represents the temperature values on a logarithm scale. The colour saturation ranges from approximately 3000~K (dark areas) to 40000~K (bright areas). We could see that, in general, temperatures are lower in the bottom of the atmosphere and get hotter when moving up and reaching the chromosphere. Interestingly, there are two distinct regions at approximately $Y=[20,40]$ arcsec and $Y=[55,65]$ arcsec where the temperatures are hotter than their surroundings at lower layers. Those regions correspond to the spatial location of the flare ribbon.

In the case of the LOS velocity (second panel from the top), there are generally low velocities at the bottom of the photosphere. Those velocities, sometimes shifted slightly towards the red, remained relatively constant with height until around $\log {\tau}=-4.5$. Only two remarkable areas deviate from that smooth behaviour and correspond to the hotter regions in the temperature panel. In this case, those areas are both dominated by strong upflows of around -20 km/s (blue colour) that extend from around $\log {\tau}=-2.5$ up to $\log {\tau}=-5.5$. Both cases show an abrupt change of sign and become strong downflowing materials above $\log {\tau}=-5.5$ (however, see below for further discussion). It is also noteworthy that velocities between those areas (along the $Y$-axis) are low, that is, close to zero. 

Regarding the field strength (third row from the top), we have a complex pattern that varies rapidly from pixel to pixel along the $Y$-axis. Generally, the stratifications are smooth with height, with no abrupt changes along the LOS for most pixels. The area corresponding to the active region, $Y=[0,20]$ arcsec, shows a larger amplitude, on the order of kilogauss, on average, while the field strength is weaker for the rest of the FOV. Additionally, the field strength decreases monotonically with height in most cases, except at around $Y=35$~arcsec, where we detected high values ($\sim$500-750 G) at upper layers (i.e. $\log \tau=-4$). The latter region corresponds to the location of the flare ribbon (see the contour line corresponding to the region of enhanced temperature at $\log \tau=-4$ in Figure~\ref{fig:atmospheric_param}).

In the case of the inclination of the magnetic field (bottom row), we have, as mentioned before, a single polarity pervading the selected spatial domain, except in the central part, where we have the opposite polarities we described as a double protrusion feature in Figure~\ref{fig:atmospheric_param}. These opposite-polarity areas are co-spatial with the low-velocity regions between the patches with upflowing material. Also, we have that, in general, the inferred vertical stratification of inclination is smooth with height.

\subsubsection{Exploring the capabilities of the inversion configuration}\label{RF_acc}

We have inferred the atmospheric parameters for all the pixels in the observed FOV during the inversion process. The results from the inversion are height-dependent (optical depth) atmospheric parameters for each pixel. However, we wanted to explore the uncertainties of the solution. In other words, we wanted to understand how reliable the atmospheric parameters are and at which optical depths they are most sensitive. 

In this regard, one option to assess this accuracy is to explore the sensitivity the spectral lines have to changes in the atmospheric parameters through the response functions (RFs). The RFs are computed by applying small perturbations, compared to the amplitude of the atmospheric parameter, at different atmospheric layers and then checking how the spectral lines are modified because of that perturbation performing a synthesis. Hence, the RFs will depend on many elements of our configuration, from the selection of the spectral lines to the properties of the observation, such as the spectral resolution, sampling, and atmospheric models. In this context, we only have one observation configuration, which simplifies the process, although the atmospheres we have inferred are very different from pixel to pixel. Thus, to assess the sensitivity and hence get information on the accuracy of the inferred parameters, we could compute the RFs for multiple pixels that belong to various physical scenarios, from the flare ribbon to regions with less activity. In that sense, we can study how the RFs differ and how large or small the amplitude of the RF is for each specific scenario and each specific optical depth. Thus, we wanted to examine the RFs for each pixel of the vertical cut crossing the flare ribbon highlighted in Figure~\ref{fig:atmospheric_param}.

\begin{figure*}
	\centering
	\includegraphics[width=18cm]{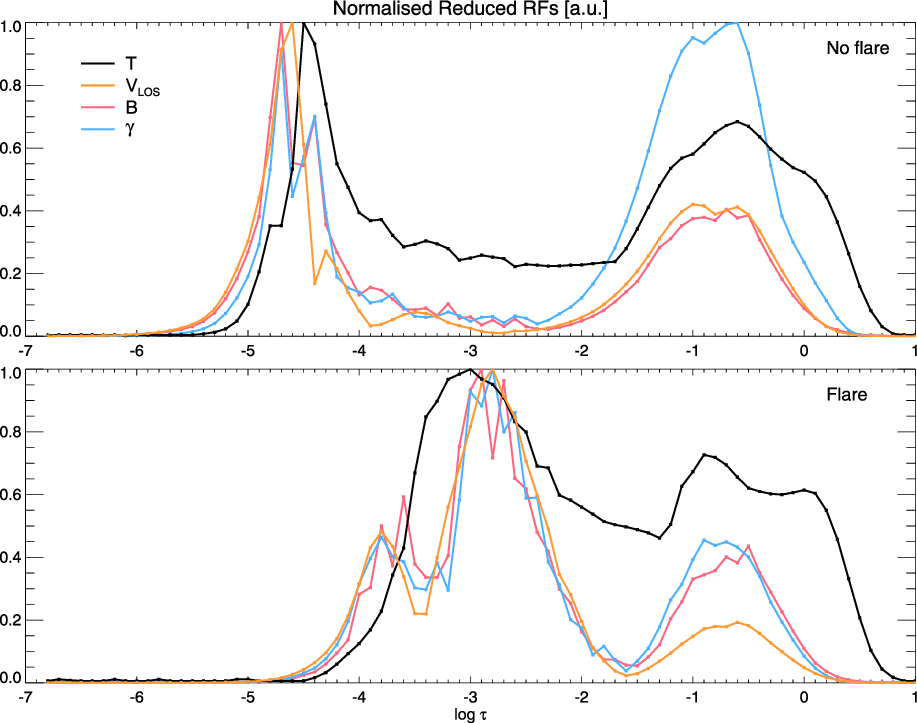}
\caption{Reduced 1D RFs to display the effective sensitivity of our observation configuration. Each panel shows the 1D RFs for temperature (black), LOS velocity (orange), magnetic field strength (red), and inclination (blue). Each line corresponds to the maximum RF value for the four Stokes parameters and all the wavelengths belonging to the \ion{Fe}{i} and \ion{Ca}{ii} spectral lines. Panels correspond to a pixel located outside the flare ribbon (top) and inside the flare ribbon (bottom).}
	\label{fig:rf_1D}
\end{figure*}

DeSIRe only computes an approximation of the NLTE RFs at the position of the nodes for each atmospheric parameter during the inversion, and as explained in \cite{DeSIRe}, the RFs are the LTE analytical RFs corrected by the NLTE departure coefficients. So in this context, we prefer to spend a greater amount of computational time and compute the RFs numerically following \cite{quinteronoda2016} for the entire optical depth range used in the inversions. In that sense, we perturb each atmospheric parameter at each optical depth and then perform the synthesis in NLTE. However, the results of this computation correspond to a large array of dimensions: $Y$, $\log {\tau}$, $\lambda$, and 4, with $Y$ being the number of spatial points (830 pixels) across the FOV, $\log {\tau}$ as the number of optical depth points in the atmosphere (81 points), $\lambda$ representing the number of wavelength samples (36 points) of our configuration (including the two spectral lines), and 4 corresponding to the number of Stokes parameters. Hence, we believe it is challenging to visualise such information without simplifying the original data. In that sense, we followed \cite{QuinteroNoda2017} by computing an equivalent 1D RF or sensitivity function. This process consisted of extracting the maximum value of the RF at a given optical depth for all the wavelength samples and all the Stokes profiles. In that sense, we collapsed the ($\log {\tau}$, $\lambda$, 4) domain into a 1D optical depth domain. Thus, we could examine where the RFs are large independently of the Stokes parameter or the spectral line, as the goal is essentially to estimate if our configuration is sensitive to changes on a given atmospheric parameter and at a given optical depth.

Figure~\ref{fig:rf_1D} shows the simplified RFs for two representative pixels. We selected one pixel belonging to the flare ribbon area (bottom) and a second one distant from it in a quieter solar area (top) in order to study the differences between atmospheric scenarios. For both pixels, we observed similarities between each atmospheric parameter, with two distinct peaks at the lower and upper layers. In general, we can assume (e.g. following \cite{QuinteroNoda2017}) that the peak in the photosphere corresponds to the contribution from continuum wavelengths, the \ion{Fe}{i} transition, and the outer wings of the \ion{Ca}{ii} line. The peak (or peaks) at upper layers corresponds to the contribution from wavelengths close to the line core of the \ion{Ca}{ii} transition. There is also a valley (a region of low sensitivity) between the peaks where the sensitivity drops, as mentioned in the works of \cite{Uitenbroek2006}, \cite{Cauzzi2008}, and \cite{delaCruzRodriguez2012} for instance. Thus, this is somehow the expected overall behaviour for this combination of spectral lines and confirms that this configuration seems capable of inferring the properties of solar phenomena from the photosphere to the lower-middle chromosphere. 

However, when comparing the results for each pixel, we noticed some differences. In particular, while the height of maximum sensitivity for the pixel outside the flare ribbon (top) is approximately where we expected it to be (i.e. around $\log \tau \sim -5.0$), the location of the peak is shifted towards deeper layers, around $\log \tau = -3$, in the case of the flare ribbon. There is still a secondary peak at $\log \tau = -4$ in the case of the inferred atmosphere inside the flare ribbon, but it is located deeper than for the standard quiet Sun atmosphere. This result is interesting, as it points out that the flare atmosphere, characterised by strong temperature enhancements plus large velocities and magnetic field concentrations, has affected the formation of the \ion{Ca}{ii} transition. Additionally, we did not see such a change in the case of the lower peaks related to the \ion{Fe}{i} spectral line. This is consistent with the fact that although the intensity profiles of \ion{Ca}{ii} are completely unusual with strong emission peaks, the profiles for the photospheric line are quite common. In other words, and in agreement with the inversion results presented in Figure~\ref{fig:atmospheric_param}, the flaring activity mainly occurs at the middle and upper atmospheric layers, with a lesser impact in the photosphere.

In addition, these results explain why we found a better spatial distribution of the atmospheric parameters on the flare ribbon area at $\log \tau = -4$ than at higher layers, where the results appear less smooth and have a salt-and-pepper pattern for most of the atmospheric parameters. The latter is usually a consequence of less sensitivity due to low values for the RFs, among other things. Thus, this study reinforces our decision to present the inversion results in Figure~\ref{fig:atmospheric_param} at $\log \tau = -4$ as a representation of the upper atmosphere where the \ion{Ca}{ii} line is sensitive instead of a more traditional value such as $\log \tau = -5$. 

Moreover, seeing this interesting behaviour, we decided to use it to better understand where our maximum sensitivity to the atmospheric parameters at higher layers is. Our strategy to do so consisted of finding the location of the maximum peak from the chromospheric contribution for the atmospheric parameters. The results for the height of that maximum peak for each atmospheric parameter and each pixel of the region highlighted with a dashed line in Figure~\ref{fig:atmospheric_param} are presented with a solid line in Figure~\ref{fig:cutx200arc}. One can see that in the areas outside the flare ribbon (e.g. $Y=[0,20]$~arcsec), the height of maximum sensitivity is, on average, at around $\log \tau \sim -5$. At the same time, it drops to lower layers in the areas where the solar flare took place. We observed a strong correlation with the regions with enhanced temperature and strong LOS velocities. This result also suggests that we have low sensitivity to changes in the atmospheric parameters at the upper layers of the flare ribbon areas. In other words, we should not trust the changes on the sign of the LOS velocity at around $\log \tau = -5.5$ for the areas $Y=[20,40]$~arcsec and $Y=[55,65]$~arcsec, as the amplitude of the RFs at those heights is very low.

%

\section{Discussion}

It is worth noticing the apparent correlation between the areas with enhanced temperature at $\log\tau = -4$ (see Figure~\ref{fig:atmospheric_param}) and the lack of velocities at the same location. Even more important is that these low-velocity areas are surrounded by areas of very large upflows, which is what we would associate with, for instance, the signature of chromospheric evaporation. In that sense, we know that the profiles for the enhanced temperature regions are very complex, with the intensity profile of the \ion{Ca}{ii} spectral line at line core wavelengths in emission, so we need to investigate further why the velocities are so low only in those areas. It is true that this region also has strong Stokes~$V$ signals (see Figure~\ref{fig:obsprof}), and they were accurately fitted. Hence, we believe strong velocities would be easily noticeable through strongly Dopplershifted circular polarisation signals, which we did not detect in the hot ribbon area. Thus, we do not have a clear answer to why velocities are so low only in the ribbon (see the contours in Figure~\ref{fig:atmospheric_param}). We plan to analyse the following snapshots from the observation to shed more light on this interesting behaviour.

We would also like to discuss the results for the magnetic field strength at higher layers (e.g. at $\log\tau = -4$). We have magnetic fields with a strength of up to 750-1000~G in the areas corresponding to the hottest part of the flare ribbon (see contours in Figure~\ref{fig:atmospheric_param}). We can also see that the magnetic field inclination in that region is almost vertical with respect to the surface, with values around 160 to 180 degrees, presenting a much smoother pattern than in other areas. We do not have a clear answer as to why the magnetic field appears so strong at chromospheric levels in that area. We believe the main contributor to the enhanced Stokes~$V$ signals in the \ion{Ca}{ii} spectral line (see Figure~\ref{fig:obsprof}) could be the enhanced intensity profiles due to the large temperatures in the flare ribbon. At the same time, it is also interesting to see that the enhanced inferred magnetic field (see Figure~\ref{fig:atmospheric_param}) seems to be correlated with the circular polarisation signals shown in Figure~\ref{fig:obsprof}. So there may be a degeneracy between the inferred temperature and the longitudinal magnetic field. This degeneracy is described, for instance, in Figure 1 of \cite{2016LRSP...13....4D}. However, our impression is the degeneracy should be smaller when analysing the full Stokes vector from two spectral lines with different properties and sensitivity to the temperature and magnetic field vector. On the other hand, it is also true that, at chromospheric levels, the only transition sensitive to the atmospheric parameters is the \ion{Ca}{ii} spectral line. Thus, at this moment, we are not able to determine the reason for this enhanced magnetic field and whether it is simply the trace of a significant temperature enhancement. We believe that with high temperatures and weak magnetic fields, it would be challenging to have such large Stokes~$V$ signals, as seen in Figure~\ref{fig:obsprof}, but at the same time, it may also be true that we are not inferring the correct values for the magnetic field strength. This is an important topic that we plan to explore further in following publications. For example, we aim to establish how much degeneracy we have between the temperature and the longitudinal field with inversions of synthetic profiles, starting from the same inversion configuration used in this work.

\section{Summary}

As far as we know, this is the first time someone has attempted to perform NLTE inversions of the full Stokes vector of a multi-line configuration observation of an X-class solar flare. Hence, the methodology itself was a research process, where we used the newly developed DeSIRe code and performed multiple tests to determine an inversion configuration that could fit this challenging data set. Particular emphasis was dedicated to fitting the complex \ion{Ca}{ii} intensity profiles that appear on emission, sometimes with values up to two times the average continuum intensity. However, the polarisation Stokes profiles were fitted simultaneously with high accuracy, although in some cases we could not reproduce the complex linear polarisation signals, mainly for the chromospheric spectral line. Still, we managed to get accurate fits, mainly on the flare ribbon area, which encouraged us to examine the results and the inferred atmospheres in detail. We detected significant temperature enhancements that mostly happen at the middle and upper layers. This could explain why the intensity profiles for the \ion{Fe}{i} spectral line are quite standard everywhere. We also detected strong upflows at the middle and upper layers, which indicates that the flaring event is producing a hot material that is moving outwards from the Sun. Interestingly, there is no indication of plasma moving down, at least not at high velocities. The information about the magnetic field vector, one of the critical novelties of this work, revealed relatively smooth stratifications with height for both strength and inclination. The Stokes profiles, mainly Stokes~$V$, do not appear to be strongly asymmetric, so we believe the smooth stratification along the LOS seems to be an accurate solution for the magnetic field vector. Still, we observed the presence of large-scale mixed polarities in the regions where the flare ribbon is located. The interaction between those mixed polarities could be the flare's triggering mechanism. We proceeded to estimate the range of heights where our configuration is sensitive to changes in the atmospheric parameters computing the RFs. We found that the spatial locations belonging to the flare ribbon present a significant depression on the height of formation (or sensitivity) for the chromospheric line. These results are consistent with those of \cite{2018Kuridze} and \cite{2021yadav}, who demonstrated, in two distinct events, that the formation height of the \ion{Ca}{ii} transitions is shifted towards deeper layers (around $\log \tau =-3$) on the spatial locations where the flare takes place. At the same time, we found no clear indication that the same happens for the \ion{Fe}{i} transition. These results again confirm that the main impact of the flaring activity happens at chromospheric levels.

Despite being able to infer the information of the solar atmosphere for a flare region through a complex inversion process, we are far from completing this research. We aim to study the time evolution of the flaring activity, performing similar inversions for multiple snapshots. At the same time, we want to better understand the limitations of our approach. A flare is probably not the best-case scenario to solve the radiative transfer equation considering hydrostatic equilibrium (HE). The truth is that we do not know how far we are from that approximation nor how large the impact of the deviation from HE on the spectral lines  will be. We performed some inversion tests without imposing HE during the computations, but the results for the fits were worse than the ones obtained with the current configuration. So we plan to test it further, mainly trying to understand if there are stark differences between the atmospheric parameters between the two assumptions when we can obtain good fits with both configurations. Nevertheless, we may conclude that the effect of departures from HE are not very important, as mentioned in \cite{2016LRSP...13....4D}.

Some researchers have highlighted the importance of NLTE effects for the \ion{Fe}{i} 617.3~nm spectral line, which may affect the accuracy of the results \citep{2023A&A...669A.144S}. In our case (i.e. in this analysis), we solved the atomic populations of the iron transition in LTE while simultaneously computing the populations of the \ion{Ca}{ii} atomic levels in NLTE. We are not sure how significant the NLTE effects on the iron spectral line are in this context, so in the future we plan to study whether there are meaningful differences when inverting the two atomic species in NLTE versus the configuration used in this work. Additionally, we want to expand this work with a thorough analysis of the implications of the atmospheres we are inferring to understand the flaring process in terms of the amount of energy transferred to upper layers or whether we have various mechanisms such as chromospheric evaporation. Most importantly, it is essential to also study the evolution of the process in order to try to understand what the trigger mechanisms are. Thus, our goal is to expand the results presented in this publication in the future. Still, for now, we believe we have taken the proper steps in this publication to analyse the data with the most advanced numerical tools and configuration we can think of, thus expanding our knowledge of solar flares a little further.

\section*{Acknowledgements}
We appreciate the helpful comments provided by B. Ruiz Cobo. The research leading to these results has received funding from the European Union’s Horizon 2020 research and innovation program under grant agreement No.739500 (PRE-EST project) and No. 824135 (SOLARNET project). F. Ferrente and F. Zuccarello acknowledge support by the Università degli Studi di Catania (PIA.CE.RI. 2020–2022 Linea 2) and by the Italian MIUR-PRIN grant 2017APKP7T on “Circumterrestrial Environment: Impact of Sun-Earth Interaction.” This research was carried out in the framework of the CAESAR (Comprehensive spAce wEather Studies for the ASPIS prototype Realization) project, supported by the Italian Space Agency and the National Institute of Astrophysics through the ASI-INAF agreement no. 2020-35-HH.0 for the development of the ASPIS (ASI Space weather InfraStructure) prototype of scientific data center for Space Weather. C. Quintero Noda acknowledges support from the Agencia Estatal de Investigación del Ministerio de Ciencia, Innovación y Universidades (MCIU/AEI) under grant ``Polarimetric Inference of Magnetic Fields'' and the European Regional Development Fund (ERDF) with reference PID2022-136563NB-I00/10.13039/501100011033. The publication is part of the Project ICTS2022-007828, funded by MICIN and the European Union NextGenerationEU/RTRP. We also thank Dr. Serena Criscuoli for providing the IBIS calibrated data.

The National Solar Observatory is operated by the Association of Universities for Research in Astronomy, Inc.(AURA) under cooperative agree- ment with the National Science Foundation. The SDO/HMI data used in this paper are courtesy of NASA/SDO and the HMI science team. Use of NASA’s Astrophysical Data System is gratefully acknowledged.\\
\textit{Facilities:} DST (IBIS), SDO (HMI)

\bibliographystyle{aa} 
\bibliography{template_new} 

\end{document}